\documentclass[11pt]{article}

\usepackage{graphicx}
\usepackage{comment}
\usepackage{authblk}

\usepackage{amsmath}
\usepackage{amssymb}
\usepackage{amsthm}
\usepackage{times}
\usepackage{color}
\usepackage{multirow}
\usepackage[mathscr]{euscript}
\usepackage{braket}
\usepackage{mathtools}
\usepackage{enumerate}

\usepackage{tikz}
\usetikzlibrary{quantikz,fit,backgrounds}
\usepackage{subcaption}       
\usepackage{booktabs}         

\usepackage{url}
\usepackage{parskip}
\usepackage{listings}
\usepackage{xcolor}

\usepackage{hyperref}
\hypersetup{
    colorlinks=true,
    linkcolor=blue,
    filecolor=blue,      
    urlcolor=blue,
    citecolor=blue
    }

\usepackage[margin=0.75in]{geometry}
\usepackage[acronym]
{glossaries}
\usepackage{nomencl}
\usepackage[font=small]{caption}
\makenomenclature

\makeglossaries

\graphicspath{ {.pdf} }
\usepackage{wrapfig}

\numberwithin{equation}{section}

\usepackage{lmodern}
\usepackage[T1]{fontenc}

\usepackage{setspace}
\usepackage{titlesec}

\begin{document}

\date{}

\title{Neuronal Spike Trains as Functional-Analytic Distributions: Representation, Analysis, and Significance}

\author{Gabriel A. Silva \\Department of Bioengineering and Department of Neurosciences  \authorcr University of California San Diego, La Jolla CA USA\\Email: gsilva@ucsd.edu}
\maketitle

\renewenvironment{abstract}
{\begin{quote}
\noindent \rule{\linewidth}{.5pt}\par{\bfseries \abstractname.}}
{\medskip\noindent \rule{\linewidth}{.5pt}
\end{quote}
}

\begin{abstract}
The action potential constitutes the digital component of the signaling dynamics of neurons. But the biophysical nature of the full-time course of the action potential associated with changes in membrane potential is mathematically distinct from its representation as a discrete set of events that encode when action potentials are triggered in a collection of spike trains. In this paper, we develop from first principles a unified functional-analytic framework for neuronal spike trains, grounded in Schwartz distribution theory. We show how this representation provides an exact operational calculus for convolution, distributional differentiation, and distributional support, which enables closed-form analysis of spike train dynamics without discretization, rate approximation, or smoothing. We then analyze the framework in the context of a two-neuron reciprocal circuit with propagation latencies and refractoriness, deriving exact results for synaptic drive, spike timing sensitivity, and causal admissibility of inputs, quantities that are either ill-defined or require approximation in conventional treatments.
\\
\end{abstract} 

\tableofcontents

\vspace{0.75in}
\section{Introduction}
A neuronal action potential, as a dynamic transient change in the membrane potential of a neuron, has the shape and kinetics it has because of the underlying biophysics of the change in membrane permeability (conductance) to (primarily) $Na^+$ and $K^+$ ion channels. The conditions responsible for the diffusion (concentration) and electrochemical driving forces that move ions across the membrane are necessarily set up \textit{before} an action potential is triggered, resulting in the instantaneous passive flux of $Na^+$ and $K^+$ ions acting as charge carriers responsible for the currents and resultant transient changes in membrane potential we call an action potential. For foundational references and background on the physiology and biophysics of the action potential, Vander's Human Physiology is particularly good at an introductory undergraduate level \cite{widmaier2023vanders}, while Kendel et al. is the standard graduate-level textbook in the field \cite{kandel2021principles}. Although there are many other superb resources, both online and in print.

Under normal physiological conditions, the shape of the neuronal action potential is stereotyped and unchanging. Every time an action potential is triggered, it proceeds to completion uninterrupted in exactly the same way with the same time course, independent of anything else the neuron is doing. As alluded to above, this is because the biophysical conditions responsible for the processes that mitigate the changes in membrane potential are set \textit{a priori}. From an information-theoretic perspective, the action potential constitutes the digital component of a neuron's signaling dynamics. The information it contributes to the system is in the timing of its occurrence, not in its shape, amplitude, or time course. 

\begin{figure}[ht] 
    \centering 
    \includegraphics[width=5in]{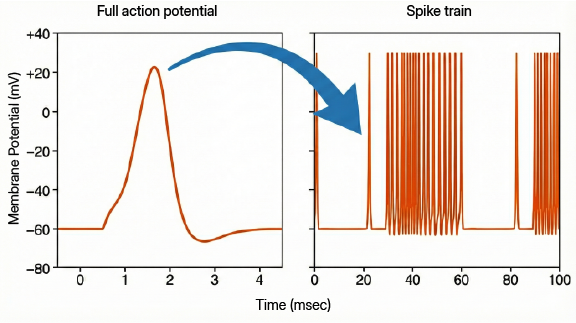} 
    \caption{The often shown but inacurate picture of an action potential, which is a stereotyped change in the membrane potential due to the underlying biophysics, compressed into a spike intended to represent it.} 
    \label{fig:ap2spikes} 
\end{figure}

It is tempting to think of a spike as a very sharp voltage waveform. A tall, narrow pulse that looks like a spike on a compressed time axis, compared to the time axis of a full action potential (Figure \ref{fig:ap2spikes}). But a spike is \textit{not} a compressed action potential, nor does it encode or represent a change in the membrane potential. In fact, a spike is not a signal in the usual sense at all. 

At the level of spiking models, what is recorded is not waveforms but events, the times at which something happened. The spike encodes the \textit{information} communicated by the action potential being triggered, not the action potential itself. For example, a neuron fires at 12.3 ms. Then again at 18.7 ms. Then at 41.2 ms. The spike train of that neuron, at its core, is just a collection of time stamps.

This immediately creates a disconnect with the mathematical machinery we typically use to describe and model physical systems. Differential equations, filters, convolutions, and dynamical systems are built to operate on functions of time, objects that assign a value to every instant. A list of isolated time points does not fit naturally into this framework. Critically, we need a way to mathematically model spike trains that maintains functional consistency. We need a description of spike trains that represents them for what they are, but in a way that supports interactions with functions and operations that model physical processes that are very different mathematically from the description of a spike train.   

\textbf{In this paper, we develop from first principles a unified functional-analytic framework for neuronal spike trains, grounded in Schwartz distribution theory \cite{botelho2025, brayman2024, aksoy2024, muscat2024, clason2020}. Spikes are represented as generalized functions defined by their action on smooth test functions rather than by pointwise values. We show that this representation is not merely a descriptive formalism, but rather, it provides an exact operational calculus for convolution, distributional differentiation, and distributional support, which enables closed-form analysis of spike train dynamics without discretization, rate approximation, or smoothing. We demonstrate the analytical power of this framework on a two-neuron reciprocal circuit with discrete signaling event propagation latencies and refractoriness, deriving exact results for synaptic drive, spike timing sensitivity, and causal admissibility of inputs, quantities that are either ill-defined or require approximation in conventional treatments.} 

The ideas and concepts we bring together and build on draw on several decades of work in theoretical and computational neuroscience and neural dynamics. Representative treatments ranging from early point-process formulations of spike trains to modern dynamical systems and network models can be found in \cite{dayan2001theoretical}, \cite{GerstnerKistler2002}, \cite{GerstnerKistlerNaudPaninski2014}, \cite{Tuckwell1988}, \cite{PerkelGersteinMoore1967}, \cite{Knight1972}, \cite{BrunelHakim1999}, and \cite{ErmentroutTerman2010}, among many others. See also the references within these cited works. 

By making the underlying mathematical structure explicit, the paper clarifies the sense in which spike trains are treated as distributions: mathematical objects defined not as functions or by their height at a single point, but by their interactions as unique events with smooth test functions that act as 'probes'. This framework provides a rigorous basis for treating spikes as mathematical objects that isolate timing and weight as the primary variables driving neural dynamics, while maintaining consistency with neurophysiologically meaningful dynamical systems. The intent is to provide a self-contained, technically rigorous reference that bridges what is often a communication barrier between the physiological and mathematical perspectives. It offers readers a rigorous foundation for engaging with and extending the extensive literature on spike-based neural dynamics.

\section{The Representational objective}
Assume neuron $i$ produces a sequence of spike times given by

$$0 < t_i^1 < t_i^2 < \cdots < t_i^k \cdots t_i^K$$

This collection, or set, of spike times for neuron $i$ is what we refer to as its spike train. (A raster plot is a matrix of spike trains, i.e., spike trains for different neurons superimposed on the same time axis.) A spike train is not naturally a classical function of the form 
$$f:\mathbb{R} \rightarrow \mathbb{R}, \text{ } t\mapsto \mathbb{R}$$ 
for two reasons. First, the events are conceptual idealizations of instantaneous events and have zero duration. Second, and critically, what matters operationally is not the value or amplitude of the spike at an individual time $t$, but how spikes contribute to or influence downstream quantities, for example, synaptic currents, counts, a kernel, a low-pass filter, or averages over a time window. The effects of spikes on these classes of functions are necessarily almost always integrals or sums.

To define a spike train, we need to define an object $s_i$ such that any physically meaningful operation we do with it can be expressed as an integral or sum paired with a test function $\varphi(t)$ that measures the effect of $s_i$ on the operation. A valid test function will need to satisfy a number of formal mathematical properties and should be of practical neurophysiological utility, like the examples from the previous paragraph. 

Formally, we want to understand, or compute, the degree to which a spike at $t_i^k$ contributes to or affects $\varphi(t_i^k)$. For multiple spikes, we want the contributions to add, such that 
\begin{equation} \label{eq:define_si}
    \quad \langle s_i,\varphi\rangle := \sum_{k}\varphi(t_i^k)
\end{equation}
Equation \ref{eq:define_si} is then the operational requirement for a spike train, where the brackets $\quad \langle \cdot,\cdot\rangle$ denote the action of the functional (spikes) on the test function. We elaborate on this paired duality notation in Section \ref{sec:define_functionals} below. It specifies a set $s_i$ paired with a test function $\varphi(t)$ such that $s_i$ is defined by how it samples and evaluates the summed effect of $\varphi(t)$ only at the specific times events (spikes) occur.

\section{Why the Dirac delta is not a classical (Lebesgue) function} \label{sec:dirac.not.f}
We will make use of the Dirac delta as the unique mathematical object that formally encodes the spike train $s_i$ to sample a test function $\varphi(t_i^k)$ in such a way that it satisfies the requirement defined by Equation \ref{eq:define_si}. To understand \textit{how} the Dirac delta does this, and why the mathematics remains consistent, one has to first deeply understand what the Dirac delta is, because it is not a function in the usual way we think of functions as maps. 

The way the Dirac delta is typically introduced in a neurophysiological context to model spikes can be misleading. So we will first formally define the Dirac delta from first principles and review some of its properties, then return to its use to encode spike trains. 

The Dirac delta $\delta$ is often introduced by associating it with three key properties. Namely:
\begin{subequations}\nonumber
    \begin{align}
        &\delta(t - a) = 0 \text{ for } t\neq a &&\text{The delta function is zero everywhere except at } t = a\\\\
        &\delta(a) = \infty &&\text{Its magnitude is infinitely large at } t = a\\\\
        &\int_{-\infty}^{\infty}\delta(t-a)\,dt = 1 &&\text{It has an area equal to unity}
    \end{align}
\end{subequations}
In other words, it is a rectangle that is infinitely high and infinitely narrow exactly and only at $t=a$. 

Almost immediately, one gets a sense that this is not a function in the ordinary sense. By an `ordinary' or classical function, we mean an integrable Lebesgue function. A function that assigns a real number to (almost) every time $t$, in a way that is well-defined pointwise and compatible with Lebesgue integration. Formally, this means that $f : \mathbb{R} \to \mathbb{R}$ such that for each $t \in \mathbb{R}$ (or for almost every $t$, not including any discontinuities), the value $f(t)$ exists as a real number. This type of function is measurable with respect to the Lebesgue measure, so the integration of the function $\displaystyle \int f(t)\,dt$ makes sense and can be defined. (There are other conditions and implications, but for our purposes here, this is sufficient to make our point.)

If we try to treat $\delta$ as a Lebesgue function, we run into an issue. If $\delta=0$ for all $t\neq a$, then $\delta$ differs from the zero function $g(t)=0$ $\forall t$, i.e., the function that equals zero for all inputs, only at a single point. But changing a function on a single point does not change its Lebesgue integral. This means that
$$\int \delta(t-a)\,dt = 0,$$
which contradicts the area $=1$ claim. There is no classical function with those properties under ordinary integration. 

The Lebesgue integral does not care about what happens at individual points. What matters is what happens on sets with nonzero measure, such as length, area, volume, etc. So for example, the interval $[0,1]$ has a Lebesgue measure $\mu([0,1]) = 1$. An interval $[a,b]$ has $\mu([a,b]) = b-a$. But any finite countable set of points, for example, $\{3,7,9,56\}$ or $\{a\}$ or $\{0\}$ itself, has a measure equal to zero, i.e., $\mu(\{a\})=0$. 

In the context of Lebesgue integration and how it is formally defined, any two functions that differ at a single point, including $\delta$ and $g(t)=0$ $\forall t$, have the same integral. To see this, consider an arbitrary function 
$$
f(t) =
\begin{cases}
M &t = a \\
0 &t \neq a
\end{cases}
$$
for $M\in\mathbb{R}$. $f(t)$ and $g(t)$ differ only at the single point $t=a$, and both have measures $\mu(f) = \mu(g) =0$. This implies that 
$$ \int f(t)dt =\int g(t)dt$$
But we know that the integral of the zero function is zero: $\displaystyle \int g(t)dt = 0$, therefore $\displaystyle \int f(t)dt = 0$.

This holds regardless of the value of $M$. If $M \rightarrow \infty$, then $\mu((f(M))=0$. In other words, values on sets of measure zero do not contribute to the integral. This contradicts the property that $\displaystyle \int \delta(t-a)dt = 1$ for any value of $(t-a) \in \mathbb{R}$. 

The way to get around this issue is by satisfying the operational requirement of Equation \ref{eq:define_si}. We will do this by formally defining the Dirac delta $\delta$ as an object called a Schwartz distribution. 

Although the Dirac delta is not an ordinary Lebesgue function and cannot be treated as one, it can be obtained as a limit of ordinary functions in the sense of Schwartz distributions. In other words, it can be approximated arbitrarily well by ordinary functions, provided the limit is understood in the context of distributions. This is the basis for the intuitive notion often used to introduce the Dirac delta. But to do this properly, we first need to understand what Schwartz distributions are. So we will hold off on showing this until Section \ref{sec:dirac.from.ordinary.f} below.

\section{Schwartz distributions}
Schwartz distributions were introduced and formalized by \href{https://www.ams.org/notices/199809/chandra.pdf}{Laurent Schwartz} (1915-2002) in the late 1940s as a new class of mathematical objects that extended functions in order to allow differentiation and Fourier analysis for singular objects like the Dirac delta \cite{botelho2025}, \cite{brayman2024}, \cite{aksoy2024}, \cite{muscat2024}, \cite{clason2020}. They are distinct from, and not to be confused with, statistical distributions, which most neuroscientists are much more familiar with. 

Formally, a Schwartz distribution $T$ is a continuous linear functional on a space of test functions. We will carefully examine each part of this definition. 

\subsection{Function versus functional} \label{sec:define_functionals}
In the most straight forward sense, a function is a map from numbers to numbers: $$f:\mathbb{R}\to\mathbb{R},\quad x\mapsto f(x)$$ (More accurately a function maps elements in a domain set to a codomain set. They can be numbers but many other things too.) A functional maps functions to numbers: $$F:\mathcal{F}\to\mathbb{R},\quad \varphi \mapsto F(\varphi)$$
where $\varphi$ represents a function. In other words, the input to a functional is itself a function.

A well-known and easy example of a functional is the definite integral. It takes a function as its input argument and returns the area under the curve defined by the function over the specified limits. For example
\begin{subequations}\nonumber
    \begin{align}
        &F(\varphi) := \int_{-\infty}^{\infty} \varphi(t)\,dt &&\text{integrated across the entire number line}\\
        &F(\varphi) := \int_{a}^{b} \varphi(t)\,dt &&\text{integrated on the closed interval } [a,b]
    \end{align}
\end{subequations}

By definition, the Dirac delta is defined as the functional that evaluates a function at a point.
\begin{equation} \label{eq:dirac.delta}
    F(\varphi) = \langle \delta(t - a), \varphi \rangle = \delta_a(\varphi) := \varphi(a)
\end{equation}
The Dirac delta is a mathematical object (a functional) that takes a test function $\varphi(t)$ and returns a number, explicitly, the value of $\varphi(t)$ at the point $a$. Notice how the pairing $\langle \delta(t - a), \varphi \rangle$ produces a number, the value of the test function at $t=a$, and satisfies the operational requirement set up in Equation \ref{eq:define_si}. This defines the Dirac delta for the single-valued case, but not (yet) a spike train $s_i$, which we work up to below.

In functional analysis, the standard notation $\langle T, \phi \rangle$ represents a duality pairing. It describes the interaction between two distinct mathematical objects: a functional $T$, in our case, the spike train, and a test function $\phi$.
Although this notation resembles an inner product, such as the dot product between two vectors, in the functional-analytic context it reflects a fundamentally different operation. Instead of multiplying two functions point-by-point, this duality notation implies an operation, the explicit application of the rule defined by $T$ to the function $\phi$.

Formally, if $T \in \mathcal{D}'(\mathbb{R})$ is a distribution and 
$\phi \in \mathcal{D}(\mathbb{R})$ is a test function, we write
\begin{equation} \nonumber
    \langle T, \phi \rangle := T(\phi)
\end{equation}
to denote the action of the continuous linear functional $T$ on $\phi$. $\mathcal{D}(\mathbb{R})$ is the space of infinitely differentiable test functions $\phi: \mathbb{R}\rightarrow \mathbb{R}$, while $\mathcal{D}'(\mathbb{R})$ is the dual space of distributions on $\mathbb{R}$. In Sections \ref{sec:test.functions.i} and \ref{sec:test.functions.ii} below we fully develop what these test functions are within the context of spike trains. 

The transition from functions to functionals reflects a fundamental shift in how a mathematical object is conceptualized. A classical function maps a unique output for every input. A functional, however, provides a measurement for every test. In this framework, the test function $\varphi(t)$ acts as a mathematical probe. There is no interaction with a distribution directly. A distribution is is characterized only by observing how it "weights" or filters different probes (test functions) that interact with it.

\subsection{Linearity of functionals}\label{sec:linearity}
Functionals also satisfy linearity, in the usual sense we think of linearity in vector spaces. Linearity must satisfy 
$$F(\alpha \varphi + \beta \psi) = \alpha F(\varphi) + \beta F(\psi) \quad \text{for all } \alpha,\beta\in\mathbb{R}$$
An integral $\displaystyle F(\varphi)=\int \varphi(t)\,dt$ is linear because it can be written as
    $$F(\alpha\varphi+\beta\psi) = \int (\alpha\varphi+\beta\psi)\,dt = \alpha\int\varphi\,dt + \beta\int\psi\,dt = \alpha F(\varphi)+\beta F(\psi)$$

Similarly, the Dirac delta is linear because it satisfies
    $$\delta_a(\alpha\varphi+\beta\psi) = (\alpha\varphi+\beta\psi)(a) = \alpha\varphi(a)+\beta\psi(a) = \alpha \delta_a(\varphi)+\beta \delta_a(\psi)$$
From the perspective of modeling spike trains using the Dirac delta, linearity is critical because it is the property that supports the superposition of spike contributions and the resultant effect on the test function under consideration.

\subsection{The space of test functions: Definition of support} \label{sec:test.functions.i}
As introduced above in Section \ref{sec:dirac.not.f}, test functions are smooth, well-behaved functions that probe a distribution. We observe the effects of the distribution on them. Schwartz distributions, like the Dirac delta, are not defined pointwise. They do not evaluate an input that maps to an output, and so only make sense in context with how they interact with test functions. 

Test functions $\varphi(t)$ are not necessarily biophysically measurable or derived variables like the membrane potential, synaptic currents, or ionic conductances. They are any smooth time-localized filter capable of interrogating a distribution. Explicitly, in our case, the distribution being constructed is a spike train encoded by Dirac deltas. We need to restrict the space of test functions to only those with these properties because, for arbitrary functions, evaluation at a point may not be defined, differentiation may be undefined, and convergence may not hold. All of which we need in the interaction between a functional and a test function. Below, we rigorously formalize these requirements.

To model a spike train, we restrict the space of test functions to $C_c^\infty$, the space of smooth functions with compact support on $\mathbb{R}$. $C$ implies continuous functions, while $C^\infty$ implies continuous functions for which all derivatives exist, i.e., are infinitely differentiable. This makes the space of test functions not just continuous but smooth, a stronger condition than continuity; all smooth functions are continuous, but not all continuous functions are smooth. Formally,
$$f \in C^\infty(\mathbb{R})
\quad \Longleftrightarrow \quad
\frac{d^k f}{dt^k}(t)\ \text{exists for all } k \in \mathbb{N}$$

The subscript in $C_c^\infty$ means the set of smooth functions that are compactly supported, meaning if $\varphi(t) \in C_c^\infty \rightarrow\varphi(t)=0$ outside some finite interval. 

The requirement for test functions to be smooth ($C^{\infty}$) is a strategic choice that shifts the burden of differentiation. Because many distributions (like the spike train $s_i$) are singular events and lack classical derivatives, distribution theory transfers the operation of differentiation onto the test function. Because $\varphi(t) $is infinitely differentiable, the definition of differentiation of the spike train is dependent on the derivatives of the test function. This ensures that the mathematical machinery never breaks down when encountering singularity associated with the events that make up a distribution like the spike train. We discuss this in some detail in Section \ref{sec:spike.diffs} below. 

Let $f:\mathbb{R}\to\mathbb{R}$ represent a classical Lebesgue function (see Section \ref{sec:dirac.not.f}). We define the support of such a function $\mathrm{supp}(f)$, as the closure of the subset of the domain on which $f$ is nonzero:
\begin{equation}
    \mathrm{supp}(f) := \overline{\{t\in\mathbb{R} : f(t)\neq 0\}}
\end{equation}
This definition relies on pointwise values. The support is defined as a subset of the function's domain (the subset where it is nonzero), and includes boundary points wherever the function is nonzero arbitrarily close to them. This means that support is not defined by where a function is nonzero at that exact point. It is defined by where it is nonzero \textit{arbitrarily close to that point}. A point in the domain of $f(t)$ belongs to the support if the function cannot be made identically zero in any arbitrarily small neighborhood around that point. This definition is designed to capture the notion of local influence, not pointwise values of $f(t)$ for some value of $t$. If this was not the case, and we attempted to define support on a pointwise basis for all nonzero values, i.e., $\{t : f(t)\neq 0\}$, then boundary points would be arbitrarily excluded, since any nonzero value of $f(t)$ could be made to approach the boundary point as close as we like. 

For example, if $f(t)=e^{-t^2}$, then $f(t)\neq 0$ for all $t$, so $\mathrm{supp}(f)=\mathbb{R}$. Given some other function $f(t)$ that is zero outside the closed interval $[a,b]$, its support is $\mathrm{supp}(f)=[a,b]$. 

The support of a distribution $T$, $\mathrm{supp}(T)$, is defined as the complement of the largest open set on which the action of $T$ on any test function supported by that set, vanishes. Formally, a point $t_0\in\mathbb{R}$ is not in $\mathrm{supp}(T)$ if and only if there exists an open neighborhood $U\ni t_0$ such that
$$\langle T,\varphi\rangle = 0 \quad \text{for all test functions }\varphi \text{ with } \mathrm{supp}(\varphi)\subset U.$$
Equivalently, $t_0\notin \mathrm{supp}(T) \quad\Longleftrightarrow\quad T\text{ is blind to everything happening near }t_0$.
Put more intuitively in the context of a test function $\varphi(t)$, if we think of a distribution as a sensor interacting with a probe defined by the test function, $\mathrm{supp}(T)$ is the set of times during which the sensor (distribution) may respond to the test. Outside of that set, it is effectively completely blind, regardless of what $\varphi(t)$ is doing.

We can now use this definition to derive the support of the Dirac delta $\delta(t-a)$. Assume that $\delta(t-a)=1$ and zero otherwise. First, we show how $\delta$ vanishes away from $a$. Take any point $t_0\neq a$. There exists an open interval $U$ containing $t_0$ but not containing $a$. Next, assume any test function $\varphi$ with $\mathrm{supp}(\varphi)\subset U$. Since $a\notin \mathrm{supp}(\varphi)$, we have $\varphi(a)=0$. By definition, though,
$$\langle \delta(t-a),\varphi\rangle = \varphi(a) = 0$$
This implies that there is an open neighborhood around every point $t_0\neq a$ on which $\delta(t-a)$ acts as zero. Therefore, $t_0\notin \mathrm{supp}(\delta(t-a))$ for all $t_0\neq a$.

In contrast, we can prove that $\delta$ does not vanish at $a$. This time, consider $t_0=a$. Take any open neighborhood $U\ni a$. We assume there exists a test function $\varphi\in C_c^\infty$ such that $\mathrm{supp}(\varphi)\subset U$, and $\varphi(a)\neq 0$. This means that $\langle \delta(t-a),\varphi\rangle = \varphi(a)\neq 0$. There is no neighborhood of $a$ on which $\delta(t-a)$ vanishes, so it must be that $a\in \mathrm{supp}(\delta(t-a))$.

Combining the two conditions then implies that $\mathrm{supp}(\delta(t-a)) = \{a\}$. $\delta(t-a)$ is completely insensitive to anything that happens outside arbitrarily small neighborhoods of time $a$.

Putting all the pieces together, we represent a spike train $s_i$ as the distribution
\begin{equation} \label{eq:si}
   s_i(t)=\sum_k \delta(t-t_i^k),
\end{equation}
where the sum is understood in the sense of distributions. Since each Dirac delta
$\delta(t-t_i^k)$ has support $\{t_i^k\}$ and the spike times are locally finite, the
support of the spike train is given by
\begin{equation}
    \mathrm{supp}(s_i)
    =
    \overline{\{t_i^k : k\in\mathbb{N}\}}.
\end{equation}
This restricts support of the spike train to only the spike times. In particular, $s_i$ is not
a function of time in the classical sense, but a distribution whose action on test
functions samples them at the discrete event times $\{t_i^k\}$.

\subsection{The space of test functions: Continuity of functionals}\label{sec:test.functions.ii}
Recall from the last section that because the space of test functions $C_c^\infty$ are infinitely differentiable and smooth, it immediately implies continuity. 

But from the perspective of functionals, continuity means that if a sequence of test functions converges smoothly and locally to a limit, then the outputs produced by the functional also converge. In contrast to the usual function continuity, this is a different meaning and definition of continuity. This is necessary to ensure stability and provide physical meaning and interpretation to modeled variables and systems. 

Given a set of test functions $\{\varphi_1, \varphi_2, \cdots \varphi_n \cdots \varphi_N\} \in C_c^\infty(\mathbb{R)}$, the set of functions converge to a limit function $\varphi$, 
\begin{subequations} \label{eq:supD}
    \begin{equation}
        \varphi_n \to \varphi \quad \Longrightarrow \quad F(\varphi_n) \to F(\varphi)
    \end{equation}
    if the functions converge uniformly:
    \begin{equation}
        \sup_t |\varphi_n(t) - \varphi(t)| \to 0
    \end{equation}
    all their derivatives converge uniformly:
    \begin{equation}
        \sup_t |D^k (\varphi_n - \varphi)(t)| \to 0
\quad \text{for all } k
    \end{equation}
\end{subequations}
and the supports remain bounded in a compact set, i.e., functions do not explode to infinity. If these conditions are met, it ensures that any $\varphi_n$ is indistinguishable from $\varphi$ even under arbitrarily local inspection. Note that 'sup' here is not 'support' but 'supremum', meaning least upper bound, in this case taken over the entire variable $t$, which is why we write $\sup_t$. Notationally, we are also using $\displaystyle D^k \equiv \frac{d^k \varphi}{dt^k}$. So writing $\sup_t |D^k (\varphi_n - \varphi)(t)|$ means the largest value ($\equiv$ the smallest possible upper bound) of all the derivatives of $(\varphi_n - \varphi)(t)$ as $t$ ranges over all real numbers, i.e., $t \in \mathbb{R}$.

In the case of the Dirac delta defined by Equation \ref{eq:dirac.delta}, where $F(\varphi) := \delta_a(\varphi) := \varphi(a)$, if $\varphi_n \rightarrow \varphi$, then $\varphi_n(a) \to \varphi(a)$ for any value $t=a$. And 
$$F(\varphi_n) = \varphi_n(a) \to \varphi(a) = F(\varphi)$$
implying that the Dirac delta is, as required, a continuous functional. 

We end this section on an important technical note. The space of test functions $\mathcal{D}(\mathbb{R})$ is equipped with the standard canonical LF-topology, i.e., limit of Fr\'echet spaces. This topology is constructed as the strict inductive limit of the spaces $\mathcal{D}_K(\mathbb{R})$
\begin{equation} \label{eq:subspaceD}
    \mathcal{D}(\mathbb{R}) = \bigcup_{K \subset \mathbb{R}} \mathcal{D}_K(\mathbb{R}) = \varinjlim \mathcal{D}_K(\mathbb{R})
\end{equation}
where each subspace $\mathcal{D}_K(\mathbb{R}) = \{ \phi \in C^\infty(\mathbb{R}) : \operatorname{supp}(\phi) \subset K \}$ consists of smooth functions supported within a compact set $K \subset \mathbb{R}$. The limit notation here is the inductive limit (or equivalently the direct limit). The space $\mathcal{D}(\mathbb{R})$ is the union of all the smaller spaces $\mathcal{D}_K(\mathbb{R})$, equipped with the strongest topology for which the inclusion maps $\mathcal{D}_K(\mathbb{R}) \hookrightarrow \mathcal{D}(\mathbb{R})$ are continuous.

As a clarification point, notationally, we use $D^k$ in Equation \ref{eq:supD} above for the $k$-th derivative operator, where $k$ is an integer. This should not be confused with the subspace $\mathcal{D}_K$ in Equation \ref{eq:subspaceD}, where $K$ represents a compact set.

Mathematically, this construction is necessary because $\mathcal{D}(\mathbb{R})$ is too ``large'' to be metrizable. It cannot be described by a single distance metric or norm that simultaneously applies over the entire infinite real line. Instead, the LF-topology stitches together a family of simpler, metrizable spaces (Fr\'echet spaces) defined on finite time intervals. 

Intuitively, this formalizes the physical constraint that experimental observations are always local. While a spike train may mathematically exist on an infinite timeline, any valid measurement that describes an action of the functional on a test function must occur within a finite observation window. Consequently, convergence in this space is defined locally, consisting of a sequence of test functions $\phi_n \to \phi$ in $\mathcal{D}(\mathbb{R})$ if and only if
\begin{enumerate}
    \item There exists a common compact support (a fixed observation window) $K$ such that $\operatorname{supp}(\phi_n) \subset K$ for all $n$, and
    \item The functions and all their derivatives converge uniformly to $\phi$ within that window.
\end{enumerate}
This formal (mathematical) characterization \cite{brayman2024, clason2020} ensures that the operations of the spike train distribution remain stable and well-defined under the limits of biophysical recording.

\section{Integral notation for the Dirac delta} \label{sec:integral.dirac}
As a notational comment, $\delta$ is often written in integral notation to express the operational definition given in Equation \ref{eq:dirac.delta}:
\begin{equation} \label{eq:integral.dirac}
    \int_{-\infty}^{\infty} \varphi(t)\,\delta(t-a)\,dt = \varphi(a) := \delta_a(\varphi)
\end{equation}
When containing a functional, this expression should not be confused with the normal integration between two (Lebesgue) functions, where we multiply the two functions point by point and add up the products over time. In the context of functionals it simply means applying a linear functional, which for the Dirac delta is defined to return $\varphi(a)$. 

While seemingly unnecessary, this notation has a number of important consequential advantages. It allows us to deal with smooth functions and spikes in a single notational framework, and maintains consistency when working with convolutions and integration by parts. We show this below.

\section{How the Dirac delta arises as a limit of ordinary functions} \label{sec:dirac.from.ordinary.f}
In this section, we return to the notion introduced in Section \ref{sec:dirac.not.f} of the Dirac delta representing an infinitely narrow, infinitely tall rectangle with a unitary area. Mathematically, this results from taking the limit of an ordinary function, but we need to use the mathematical machinery of distributions and test functions to do so properly. 

We beginning by letting $\rho(t)\ge 0$ be a smooth function with:
$$\int_{-\infty}^\infty \rho(t)\,dt = 1.$$

We then define a scaled family of related functions by
$$\rho_\varepsilon(t) := \frac{1}{\varepsilon}\rho\!\left(\frac{t}{\varepsilon}\right).$$
for the parameter $\epsilon$.

Notice how the area remains equal to 1, since     
    $$\int \rho_\varepsilon(t)\,dt = \int \frac{1}{\varepsilon}\rho\!\left(\frac{t}{\varepsilon}\right) dt = \int \rho(u)\,du = 1 \text{ where, } u=\frac{t}{\epsilon}$$
The functions progressively cluster around $t=0$ as $\varepsilon\to 0$. To center the functions at the point $a$, we define $\rho_\varepsilon(t-a)$.

For any test function $\varphi\in C_c^\infty$, we can write
$$\int \varphi(t)\rho_\varepsilon(t-a)\,dt.$$
and substituting for a shifted center point $t=a+\varepsilon u$, this becomes
$$\int \varphi(a+\varepsilon u)\rho(u)\,du.$$

As $\varepsilon\to 0$, $\varphi(a+\varepsilon u)\to \varphi(a)$ pointwise, and because $\varphi$ is smooth and compactly supported (see Sections \ref{sec:test.functions.i} and \ref{sec:test.functions.ii}), convergence applies, giving
$$\lim_{\varepsilon\to 0}\int \varphi(t)\rho_\varepsilon(t-a)\,dt = \int \varphi(a)\rho(u)\,du = \varphi(a)\int \rho(u)\,du = \varphi(a).$$

So $\rho_\varepsilon(t-a)$ converges to $\delta_a$ as distributions:
\begin{equation} \label{eq:dual}
    \rho_\varepsilon(t-a)\xrightarrow{\ \mathcal{D}'\ }\delta_a
\end{equation}
Following standard functional analysis notation, we again write $\mathcal{D}'$ to indicate the space of (Schwartz) distributions. The space of distributions $\mathcal{D'}(\mathbb{R})$ is the dual space to the space of test functions: $\mathcal{D}(\mathbb{R}) = C_c^\infty(\mathbb{R})$. $C_c^\infty(\mathbb{R})$ is formally a vector space, and the dual space of distributions is the space of linear functionals. So writing $\rho_\varepsilon(t-a)\xrightarrow{\ \mathcal{D}'\ }\delta_a$ in Equation \ref{eq:dual} implies that in the limit the functions converge in the sense of distributions.

This is the meaning behind the idea of a Dirac delta as an infinitely tall, infinitely narrow, rectangle with area equal to one.

\section{Putting it all together: The dynamics of spike trains} \label{sec:all.together}
In this section, we apply the framework developed above to the dynamics of spike trains. Fundamentally, the purpose of this formalism is not just to represent spikes more accurately, but to enable mathematically rigorous and consistent modeling and analysis of neuronal dynamics that cannot be treated coherently without it. In Section \ref{sec:all.together} here, we derive some theoretical results. In Section \ref{sec:2.neuron.circuit} below, we apply these results to the analysis of a simple reciprical two neuron circuit. 

\subsection{Linear summation of spike distributions} \label{sec:linearity2}
Recall from Equation \ref{eq:dirac.delta} that $\langle \delta(t - a), \varphi \rangle = \delta_a(\varphi) := \varphi(a)$ Given that distributions are linear functionals (Section \ref{sec:linearity}), if $T_1$ and $T_2$ are distributions, then $T_1+T_2$ is defined by
\begin{equation}
    \langle T_1+T_2,\varphi\rangle = \langle T_1,\varphi\rangle + \langle T_2,\varphi\rangle
\end{equation}
\textit{c.f.} also Equation \ref{eq:define_si}.
In Equation \ref{eq:si} we defined a spike train as
\begin{subequations} \nonumber
    \begin{align}
        &s_i(t)=\sum_k \delta(t-t_i^k) &\text{Equation \ref{eq:si}} 
    \end{align}
\end{subequations}
This is not a function, it is a sum of linear functionals. Due to linearity, the action of this sum on a test function $\varphi\in C_c^\infty$ is
\begin{equation} \label{eq:sum.delta}
    \left\langle \sum_k \delta(t-t_i^k),\varphi\right\rangle = \sum_k \langle \delta(t-t_i^k),\varphi\rangle = \sum_k \varphi(t_i^k)
\end{equation}
Equation \ref{eq:sum.delta} is interpreted as sampling the test function at every spike and adding up the results, which is the intended behavior. 
 
Whereas, Equation \ref{eq:dirac.delta} satisfies the operational requirement we set up in Equation \ref{eq:define_si} for the single value case, Equation \ref{eq:sum.delta} equivalently does so for a linear combination of functionals.

\subsection{Local finiteness and well-defined spike trains} \label{sec:local.fitnite}
We next address a technical consideration to ensure that the spike train representation is mathematically well-defined. Because distributions are defined by their action on compactly supported test functions, the only way an infinite sum of spike events could fail to be well-defined is if infinitely many spikes contribute within a finite time window. This condition is avoided by assuming local finiteness, that is, only finitely many spikes may occur in any bounded interval of time, even though the spike train may contain infinitely many events overall, mathematically speaking at least. In practice, of course, any spike train is necessarily finite. But this requirement ensures mathematical consistency. 

Intentionally, test functions $\varphi \in C_c^\infty(\mathbb{R})$ have compact support (\textit{c.f.} Sections \ref{sec:test.functions.i} and \ref{sec:test.functions.ii}). This ensures that there exists a finite interval $[a,b]$ such that $\varphi(t)=0$ outside $[a,b]$.
Because $t\in \mathbb{R}$, the support of $\varphi$ takes the infinite sum 
$$\sum_k^{\infty} \varphi(t_i^k)$$ 
and restricts it to
$$\sum_{k:\,t_i^k \in [a,b]} \varphi(t_i^k)$$
since every term $t_i^k \notin [a,b]$ is zero. The only remaining condition we need to impose is that there not be infinitely many spike times inside $[a,b]$. So we assume local finiteness such that
$$\#\{k : t_i^k \in [a,b]\} < \infty \quad \text{for every bounded interval } [a,b]$$
This guarantees that for every test function only finitely many nonzero terms (spikes) appear in the sum, making the sum finite and precise. Otherwise, the sum becomes a conditionally convergent (or divergent) infinite series due to Riemann rearrangement, whereby any conditionally convergent infinite series of real numbers can be rearranged to converge to any value, or to diverge. From a neurophysiological and neural dynamics perspective, this constraint is, of course, reasonable, since we would never record an infinite number of spikes in a finite time window.

\subsection{Convolution and spike train dynamics}
In this section, we show how the spike-train representation naturally gives rise to the standard convolution dynamical description of synaptic and membrane processes. Once spikes are modeled as distributions, i.e., instantaneous events defined only through their action on test functions, dynamics is no longer a pointwise evaluation in time, but linear filtering via convolution. Each spike produces a time-shifted copy of a fixed response kernel, and multiple spikes then combine by superposition. This formalism naturally explains why synaptic dynamics take the form of shifted impulse responses, and provides the mathematical machinery to study it. We will use this example of synaptic dynamics to develop the intuition. 

We want to construct a spike train $s_i$ such that each spike at time $t_i^k$ produces a stereotyped postsynaptic response waveform kernel $K$ that gets triggered at that time. Consistent with the formalism we laid out for Equation \ref{eq:define_si}, the subscript $i$ indexes individual neurons, while the superscript $k$ indexes individual spike events within the spike train of neuron $i$. Leveraging the linear summation of multiple occurring spiking events, we can write an expression for synaptic current $I_{syn}$ as 
\begin{equation} \label{eq:Isyn}
    I_{\text{syn}}(t) = \sum_k K(t - t_i^k)
\end{equation}
Neurophysiologically, $K$ here represents the elementary postsynaptic current contributed by each presynaptic neuron on the postsynaptic neuron $i$. $I_{syn}(t)$ is the linear superposition of individual currents, i.e., the total synaptic current. For example, a standard form for a postsynaptic current kernel for ionotropic $Cl^-$ gamma-aminobutyric acid (GABA) and $Na^+$ $\alpha$-amino-3-hydroxy-5-methyl-4-isoxazolepropionic acid (AMPA) ion channels is 
\begin{equation} \label{eq:eg.kernel}
    K(t)=g_{\max}\,\bigl(e^{-t/\tau_d} - e^{-t/\tau_r}\bigr) H(t)
\end{equation}
where $\tau_{r}$ and $\tau_d$ are the waveform rise and decay time constants, $g_{max}$ is the maximum or peak channel conductance, and $H(t)$ is the Heaviside step function which acts as a switch \cite{dayan2001theoretical}, \cite{destexhe1994efficient}, \cite{destexhe1998kinetic}.

By modeling the spike train as a set of Dirac deltas $\displaystyle s_i(t) = \sum_k \delta(t - t_i^k)$, the convolution of the kernel with the delta spike train representation computes to our expression of $I_{syn}$ in Equation \ref{eq:Isyn}. 

To see this, we define convolution for an ordinary function $f$ with a kernel $K$ in the usual way as  
\begin{equation} \label{eq:kernel}
    (K*f)(t) := \int_{-\infty}^{\infty} K(t-\tau)\, f(\tau)\, d\tau
\end{equation}
For a fixed (current) time $t$, we take into account all past input times $\tau$. For each $f(\tau)$ we weight it by how much it contributes to the output at the computed time $t$ using the kernel $K(t-\tau)$. Finally, we sum (integrate) over all time points $\tau$. This is a linear time invariant filter. 

For a single spike at $\tau=a$ represented as a Dirac delta, $f(\tau)=\delta(\tau-a)$, substituting into Equation \ref{eq:kernel} yields
\begin{subequations}
\begin{equation} \label{eq:kernel2}
    (K*\delta(\tau-a))(t)=\int_{-\infty}^{\infty} K(t-\tau)\,\delta(\tau-a)\, d\tau.
\end{equation}
Recall from Equation \ref{eq:integral.dirac} in Section \ref{sec:integral.dirac} that 
\begin{equation} \nonumber
    \int_{-\infty}^{\infty} \varphi(\tau)\,\delta(\tau-a)\,d\tau = \varphi(a) 
\end{equation}
But here we have an explicit form for $\varphi(t):=K(t-\tau)$ so putting this back into Equation \ref{eq:kernel2} gives
\begin{equation}\label{eq:K.onespike}
    (K*\delta(\tau-a))(t)=\int_{-\infty}^{\infty} K(t-\tau)\,\delta(\tau-a)\, d\tau = K(t-a)
\end{equation}
\end{subequations}
A spike at $\tau=a$ results in a postsynaptic response shifted to start at $\tau=a$.

Extending this to a series of spikes, i.e., a spike train $\displaystyle s_i(\tau) = \sum_k \delta(\tau - t_i^k)$, assuming linearity (Sections \ref{sec:linearity} and \ref{sec:linearity2}) and local finiteness (Section \ref{sec:local.fitnite}), we compute the convolution in the same way (dropping the integration limits for notational simplicity): 
\begin{equation} \label{eq:K.multispikes}
    (K*s_i)(t)=\int K(t-\tau)\, s_i(\tau)\, d\tau=\int K(t-\tau)\left(\sum_k \delta(\tau-t_i^k)\right)d\tau
\end{equation}
Notice how each term in Equation \ref{eq:K.multispikes} is 
$$\int K(t-\tau)\,\delta(\tau-t_i^k)\,d\tau = K(t-t_i^k)$$
which is the multi-spike version of Equation \ref{eq:K.onespike} we derived above. This implies that 
\begin{equation} \label{eq:K.allspikes}
    (K*s_i)(t)=\sum_k K(t-t_i^k)
\end{equation}
Equation \ref{eq:K.allspikes} is the explicit interpretation of convolution in the context of spike train dynamics. Each presynaptic spike at time $t_i^k$ contributes a time-shifted copy of the elementary postsynaptic response $K$, and the total synaptic current at time $t$ is obtained by summing these contributions. Importantly, the process does not rely on treating spikes as functions of time; it follows directly from representing spikes as distributions and applying linear filtering. Convolution emerges as the unique operation that maps discrete spike events to continuous synaptic dynamics that preserves linearity, causality, and temporal locality.

In the context of convolution, the Dirac delta acts as a form of time translation. When a synaptic kernel $K$ is convolved with a time-shifted delta $\delta(t-a)$, the delta essentially translates the entire shape of the kernel along the timeline to exactly $t=a$. This functionally transforms a complex integral into a simple copy-paste operation, in the sense that for every spike time in the collection of events, the distributional operation pastes a copy of the synaptic response into the total current, allowing discrete events to be encoded into continuous-time dynamics.

This formalism shows why convolution provides the natural dynamical description for event-driven spiking neuronal processes. Once spikes are modeled as instantaneous events in the sense of distributions, linear filtering by convolution is the only operation able to encode discrete spike times into continuous-time responses. Synaptic currents, membrane potentials, and other downstream variables inherit their temporal structure directly from the kernel $K$ and the geometry of spike timing.

\subsection{Distributional differentiation of spike trains} \label{sec:spike.diffs}
In this section we discuss another property that emphasizes the unique gains of a distribution theory formalism to neuronal spike trains and important topics in neuroscience. Although a spike train has no pointwise derivative, its derivative is well-defined in the distributional sense. By definition, the derivative $s_i'$ acts on a test function $\varphi$ as $\langle s_i',\varphi\rangle = -\langle s_i,\varphi'\rangle$, producing $-\sum_k \varphi'(t_i^k)$. In other words, differentiation of a spike train does not measure changes in the spikes themselves, which has no meaning, but rather local changes in the test function at the event times. It tells us how sensitive a downstream process is to the precise timing of spikes. 

This is directly relevant to \textbf{temporal coding hypotheses}, where information is carried not by average firing rates but by the precise times at which individual spikes occur. Under a rate code, small shifts in spike timing are noise. But under a temporal code, they are signal. The distributional derivative $\langle s_i', \varphi \rangle = -\sum_k \varphi'(t_i^k)$ provides a mathematically exact way to quantify this distinction. It measures how much a given downstream process $\varphi$ changes per unit shift in spike time, without approximation or binning. Temporal coding and its contrast with rate coding have been extensively studied in sensory systems \cite{GerstnerKistlerNaudPaninski2014, theunissen1995temporal, panzeri2010sensory}.

This formalism also provides a natural framework for understanding how \textbf{spike timing jitter affects synaptic integration}. In any real neural circuit, spike times are variable, they fluctuate trial to trial due to channel noise, synaptic stochasticity, and network dynamics. The question of whether downstream computation is robust or fragile to this jitter is ultimately a question about the magnitude of $|\varphi'(t_i^k)|$ in the sense that if the downstream process changes slowly near the spike time, the computation is robust to small timing perturbations. But if it changes quickly, the computation is necessarily sensitive. This connects the distributional derivative directly to the experimental literature on spike timing precision and its functional consequences \cite{mainen1995reliability, koch1997relationship, galan2008optimal}.

Distributional differentiation also provides the formal mathematical basis for analyzing \textbf{spike-timing-dependent plasticity (STDP)}, in which the sign and magnitude of synaptic modification depend on the precise temporal order and interval between pre- and postsynaptic spikes, typically within windows of tens of milliseconds. STDP is inherently a derivative-like operation on spike timing. The plasticity rule assigns different outcomes depending on whether a presynaptic spike leads or lags a postsynaptic spike by a small temporal offset. The distributional framework developed here provides the rigorous mathematical structure within which such timing-dependent operations on event-structured objects can be formally defined and analyzed \cite{bi1998synaptic, markram1997regulation, feldman2012spike}.

Finally, the distributional derivative supports the analysis of \textbf{delay-dependent synchronization and refractoriness} in neural circuits. In networks where action potentials (or equivalent discrete signals) propagate with finite conduction velocities and neurons experience a refractory period after firing, the causal admissibility of an input depends on the precise timing relationship between signal arrival and the neuron's refractory state \cite{sila2019, puppo2018}. The derivative of a spike train, acting on a downstream kernel, quantifies exactly how perturbations in these timing relationships propagate through a circuit, a question that cannot be addressed or modeled within a rate-coding or binned-time framework \cite{BrunelHakim1999, ErmentroutTerman2010}.

We will develop the intuition of distributional differentiation carefully. 

A spike train is a set of instantaneous events, so what does it even mean to differentiate a set like that? When we differentiate ordinary functions, derivatives have physical and geometric interpretations, such as the rate of change, or slope. But distributions like the spike train are not ordinary functions. They have no meaning or values between events, or concepts like slope. This necessitates a change in perspective that ultimately results in a powerful set of mathematical ideas and tools to analyze and understand (in our case) very real world physical and physiological properties of neural signaling and dynamics. 

This shift in perspective requires abandoning the question of how to directly differentiate the distribution. Because the spike train has no pointwise values, it is not directly differentiable. Instead, we differentiate the test functions and ask how does the spike train respond. 

From the operational requirement formalized in Equation \ref{eq:define_si}, 
$$\langle s_i, \varphi \rangle = \sum_k \varphi(t_i^k)$$ 
with $s_i(t)=\sum_k \delta(t-t_i^k)$ (Equation \ref{eq:si}), the spike train samples the test function at the event times. 

For any distribution $T$, its derivative $T'$ is defined as 
\begin{equation} \label{eq:s'.defined}
    \langle T',\varphi \rangle := -\langle T,\varphi'\rangle
\end{equation}
This definition is consistent with the ordinary notion of differentiation if $T$ were a smooth function, and preserves integration by parts, but avoids pointwise evaluation. We look at this more deeply (and why the negative sign in the definition needs to be there) below. 

For a spike train, $T=s_i$
\begin{subequations} \label{eq:s'}
    \begin{align}
        \langle s_i',\varphi \rangle &= -\langle s_i,\varphi' \rangle \\
        \langle s_i',\varphi \rangle &= -\sum_k \varphi'(t_i^k)
    \end{align}
\end{subequations}
The interpretation of Equation \ref{eq:s'} is that the original spike train $s_i$ samples values of the test function $\varphi(t)$ at the event times, while the derivative of the spike train, $s_i'$, samples values of the derivative of the test function $\varphi'(t)$. In other words, $s_i'$ samples values of the slope of $\varphi(t)$. This means that the derivatives of the spike train (by definition) measures how the test function \textit{changes} at the event times. 

The implication is that $s_i'$ encodes \textit{how sensitive the test function is to timing shifts}, i.e., at the spike times. So for example, if $\varphi(t)$ has small amplitudes or is zero near some spike time $t_i^k$ then $\varphi'(t_i^k) \rightarrow 0$ or will be equal to zero at that event time, so the spike (sampling the value $\varphi'(t_i^k)$) will produce no response in its derivative. It will not contribute much or anything to the sum in Equation \ref{eq:s'}. If, on the other hand, $\varphi(t)$ changes quickly or steeply near $t_i^k$, then $|\varphi'(t_i^k)|$ will have a large value and the spike at $t_i^k$ will produce a large contribution to the sum.

Two closing technical comments: First, if $\varphi(t)$ is not differentiable, then $\varphi'(t)$ would be undefined, making $s_i'$ undefined (directly due to its definition). So this operation necessitates all the work we did above to justify why $\varphi(t) \in C_c^{\infty}$.

Second, the minus sign in Equation \ref{eq:s'.defined} results directly from integration by parts. It is worth considering why it arises in some detail, because it provides a deep intuitive insight. 

Assume some function $f(t) \in C^{\infty}$ is an ordinary smooth function that is smooth but not necessarily nonzero everywhere and possibly grows to infinity. We still require a test function $\varphi(t) \in C_c^{\infty}$, i.e., smooth \textit{and} compactly supported. We can choose to view $f(t)$ as a distribution by defining 
\begin{equation} \label{eq:fasT}
  T_f(\varphi) := \int f(t)\,\varphi(t)\,dt  
\end{equation}
Then, taking Equation \ref{eq:s'.defined} into consideration, the distributional derivative of $T_f(\varphi)$ is \begin{equation} \label{eq:fasT2}
    \langle T_f', \varphi \rangle := -\langle T_f, \varphi' \rangle = - \int f(t)\varphi'(t)\,dt
\end{equation}
Because $f(t)$ is an ordinary smooth function, and $\varphi(t)$ is smooth and compactly supported, we can integrate by parts. First, notice that due to the product rule, the derivative of the product of $f(t)\varphi(t)$ is
$$\frac{d}{dt}\bigl(f(t)\varphi(t)\bigr)=f'(t)\varphi(t) + f(t)\varphi'(t)$$
Integrating both sides of this expression over $\mathbb{R}$ produces
\begin{subequations}
\begin{equation} \label{eq:integrarte.fphi}
    \int_{-\infty}^{\infty}\frac{d}{dt}\bigl(f(t)\varphi(t)\bigr)\,dt=\int_{-\infty}^{\infty}\Bigl(f'(t)\varphi(t) + f(t)\varphi'(t)\Bigr)\,dt
\end{equation}
If we consider the left hand side of Equation \ref{eq:integrarte.fphi}, the fundamental theorem of calculus results in an evaluation of the boundary conditions 
$$\int_{-\infty}^{\infty}\frac{d}{dt}\bigl(f(t)\varphi(t)\bigr)\,dt=\Bigl[f(t)\varphi(t)\Bigr]_{t=-\infty}^{t=\infty}$$
Substituting back into Equation \ref{eq:integrarte.fphi} produces
\begin{equation} \label{eq:integrarte.fphi.2}
    \Bigl[f(t)\varphi(t)\Bigr]_{-\infty}^{\infty}=\int_{-\infty}^{\infty} f'(t)\varphi(t)\,dt+\int_{-\infty}^{\infty}f(t)\varphi'(t)\,dt
\end{equation}
Evaluate the boundaries at $-\infty$ and $\infty$. Since $\varphi(t) \in C_c^\infty(\mathbb{R})$, there has to exist some number $N_+$ such that $\varphi(t)=0$ for all $|t| > N_+$. Similarly, there must exist some $N_-$ such that $\varphi(t)=0$ for all $|t| < N_-$. This means that in the limit at $t \rightarrow \pm \infty$ the boundary term is
$$\Bigl[f(t)\varphi(t)\Bigr]_{-\infty}^{\infty}\lim_{t\to\infty} f(t)\varphi(t) - \lim_{t\to-\infty} f(t)\varphi(t)=0-0=0$$
As a result, Equation \ref{eq:integrarte.fphi.2} becomes
\begin{equation}
    0=\int_{-\infty}^{\infty} f'(t)\varphi(t)\,dt+\int_{-\infty}^{\infty} f(t)\varphi'(t)\,dt 
\end{equation}
which is equivalent to
\begin{equation} \label{eq:integrarte.fphi.3}
    -\int_{-\infty}^{\infty} f(t)\varphi'(t)\,dt=\int_{-\infty}^{\infty} f'(t)\varphi(t)\,dt
\end{equation}
\end{subequations}
Notice how the right hand side of Equation \ref{eq:integrarte.fphi.3} is (by definition) $\langle T'_{f}, \varphi \rangle$ (\textit{c.f.} Equation \ref{eq:fasT}). The left hand side is Equation \ref{eq:fasT2}: $\langle T_f, \varphi' \rangle$. Therefore, 
\begin{equation}
    -\langle T_f, \varphi' \rangle = \langle T'_{f}, \varphi \rangle \quad \text{for all } \varphi(t)
\end{equation}

The distributional derivative is a framework that extends differentiation in such a way that it integrates both smooth functions and instantaneous events without (mathematical) contradiction. It provides a unified definition of differentiation that not only applies to smooth functions, but also extends, consistently and rigorously, to sets of events where classical differentiation is not defined. Integration by parts starts with a theorem about functions and ends with a defining identity that holds for all distributions.

From a neuroscientific context, including the important neurophysiological topics discussed at the beginning of the section, the integration by parts derivation is not just a technical exercise. It is the mathematically rigorous process that ensures the distributional derivative of a spike train is consistent with the classical derivative of any smooth approximation to that spike train. This means that one can freely pass between continuous descriptions of neural dynamics (e.g., membrane potentials, synaptic conductances, dendritic integration) and event-based descriptions (e.g., spike trains, refractory resets, threshold crossings) without introducing mathematical inconsistencies. The two descriptions are not competing frameworks requiring reconciliation, but rather, projections of a single, unified structure.

This consistency has concrete analytical consequences. Because the distributional derivative $\langle s_i', \varphi \rangle = -\sum_k \varphi'(t_i^k)$ is well-defined for any smooth test function, it can be applied to synaptic kernels, membrane filters, or any other physiologically meaningful process to quantify timing sensitivity at event times. It can be composed with convolution (Section 7.3) to analyze how perturbations in spike timing at one neuron propagate through successive stages of synaptic transmission to affect downstream circuits. This process can also be extended to higher-order derivatives without additional assumptions, since the definition is recursive: $\langle T^{(n)}, \varphi \rangle = (-1)^n \langle T, \varphi^{(n)} \rangle$. This provides the formal foundation for rigorous analysis of neural phenomena such as timing-dependent plasticity, refractory gating, and delay-dependent synchronization, for which the precise timing of individual spikes, rather than average firing rates, determines the computational outcome.

\section{An applied example: Two reciprocally connected neurons} \label{sec:2.neuron.circuit}
While the preceding sections developed the concepts and mathematical machinery of a spike train distribution theory in full generality, in this last section we illustrate a concrete applied example that, while intentionally minimal, is nonetheless nontrivial and cannot be properly or rigorously modeled without the expressive analytic power of the framework. This example is the smallest system in which signal propagation latencies, synaptic filtering, timing sensitivity, and causal admissibility of signals all interact. It illustrates, in explicit detail, what the
distributional representation enables that informal treatments of spike trains cannot
provide. As we work through this example, we will reference back to the relevant sections and equations that developed the core mathematics. 

A full dynamical analysis of the circuit is beyond the scope of this paper; however, in each case, we explicitly show the mathematical setup and its analytic value, from which subsequent dynamical analysis of physiological and clinical relevance can be explored. 

\subsection{A two-neuron circuit}

Conceptually following the approach in \cite{sila2019}, consider two neurons, $A$ and $B$, reciprocally connected with each other. Neuron $A$ sends a directed
connection to neuron $B$, and neuron $B$ sends a directed connection back to neuron $A$.
Each connection has an associated temporal latency (propagation delay) given by $\tau_{AB} > 0$ for the connection
from $A$ to $B$, and $\tau_{BA} > 0$ for the connection from $B$ to $A$. These latencies are the
physical consequences of finite conduction velocities (signal propagation speeds) along the connecting pathways. Note that these pathways need not be geodesics (shortest physical distances) between the two neurons, and anatomically essentially never are. They reflect the actual convoluted path and length of reciprocally connected axons. Discrete signaling events such as action potentials traveling at finite conduction velocities propagating on pathways with some given length or distance is then what produces signaling latencies. 

Each neuron also has a refractory period following an action potential (spikes), a time interval during
which the neuron cannot respond to incoming inputs regardless of its strength. Let
$R_A > 0$ and $R_B > 0$ denote the duration of the refractory periods for neurons $A$ and $B$, respectively.

We represent the spike trains of the two neurons as distributions, following the
formalism of Sections 2 and 4 (\textit{c.f.} Equations \ref{eq:define_si} and \ref{eq:si})
\begin{equation}
s_A(t) = \sum_m \delta(t - t_A^m)  \qquad s_B(t) = \sum_n \delta(t - t_B^n)
\label{eq:two_neuron_spike_trains}
\end{equation}
where $\{t_A^m\}$ and $\{t_B^n\}$ are the spike times of neurons $A$ and $B$, respectively, i.e., the times at which each neuron sends out discrete signaling events. 
Both spike trains are distributions in the sense of Section 4, in that they are continuous
linear functionals on the space of test functions $C_c^\infty(\mathbb{R})$, and their
supports are the discrete sets of spike times:
\begin{equation} \nonumber
\mathrm{supp}(s_A) = \{t_A^m : m \in \mathbb{N}\} \qquad
\mathrm{supp}(s_B) = \{t_B^n : n \in \mathbb{N}\}
\end{equation}

Each synapse is characterized by a postsynaptic response kernel $K(t)$, as introduced
in Section 7.3. For concreteness, we can consider the standard bi-exponential form given by Equation \ref{eq:eg.kernel}:%
\begin{equation} \nonumber
K(t) = g_{\max}\left(e^{-t/\tau_d} - e^{-t/\tau_r}\right) H(t)
\end{equation}
where $\tau_r$ and $\tau_d$ are the rise and decay time constants, $g_{\max}$ is the peak
conductance, and $H(t)$ is the Heaviside step function ensuring causality, i.e., $K(t) = 0$
for $t < 0$.

Given this setup, we can then apply the three
main results of Section 7, convolution, distributional differentiation, and
distributional support, to analyze this circuit.

\subsection{Synaptic drive via convolution} \label{sec:synaptic.drive}

The first question we address is what is the total synaptic drive arriving at each neuron? This is the quantity that ultimately determines whether a neuron fires. Computing it exactly without discretizing time or approximating spike trains as rates is essential whenever the precise temporal alignment of converging inputs matters for the circuit's function. Physiologically, this precise temporal structure underlies everything from sensory processing to motor coordination. Clinically, its disruption is implicated in pathologies ranging from epilepsy, where excessive synchronous synaptic drive triggers seizures, to schizophrenia, where altered timing of excitatory and inhibitory currents degrades cortical circuit function. 

When neuron $B$ receives spikes from neuron $A$, each event arrives at $B$ with a latency $\tau_{AB}$. The delayed spike train arriving at $B$ is given by 
\begin{equation}
s_A(t - \tau_{AB}) = \sum_m \delta\big(t - (t_A^m + \tau_{AB})\big)
\label{eq:delayed_train}
\end{equation}
This is a time-shifted distribution of $s_A$, i.e., the spike train of neuron $A$ shifted forward in time by the propagation delay, with support at the
set of arrival times $\{t_A^m + \tau_{AB}\}$.

From the convolution result derived in Equation \ref{eq:K.allspikes} in Section 7.3, the synaptic
current produced at neuron $B$ by the spike train of neuron $A$ is
\begin{equation}
I_{A \to B}(t) = \int K(t-\tau)\, s_A(\tau - \tau_{AB})\, d\tau
= \sum_m K\big(t - t_A^m - \tau_{AB}\big)
\label{eq:synaptic_drive_B}
\end{equation}
Each spike from neuron $A$, emitted at time $t_A^m$, contributes a time-shifted copy of
the kernel $K$ that begins at time $t_A^m + \tau_{AB}$. The total synaptic current at $B$
is the linear superposition of these contributions. This is an exact result and not an approximation, there is no
discretization, binning, or smoothing of the spike train needed in the modeling or computation.

By the same reasoning, the synaptic drive at neuron $A$ from neuron $B$ is:
\begin{equation}
I_{B \to A}(t) = \sum_n K\big(t - t_B^n - \tau_{BA}\big)
\label{eq:synaptic_drive_A}
\end{equation}

These expressions are the direct consequences of treating spike trains as distributions
and applying convolution in the sense of distributions. They are exact, continuous-time
descriptions of the synaptic currents, valid for arbitrary (locally finite) spike
patterns. Because these expressions are exact and carry no temporal discretization, they preserve timing relationships at arbitrary precision, including the sub-millisecond regime where the relative arrival times of converging inputs determine whether a postsynaptic neuron integrates them cooperatively or independently. This distinction is critical in any setting where coincidence detection is functionally relevant. For example, in binaural auditory processing, interaural time differences of tens of microseconds have to be resolved to localize sound sources \cite{brand2002precise}. The detection of synchronous excitatory events that trigger epileptiform activity when inhibitory timing is disrupted also depends on such precise timing \cite{jiruska2013synchronization}. Rate-based or time-binned approximations necessarily sacrifice this precision, imposing a temporal resolution floor that may obscure the very timing relationships that govern circuit behavior. The distributional formulation avoids this compromise entirely.

\subsection{Spike timing sensitivity via the distributional derivative} \label{sec:spke.sensitivity}
The second question we consider is how sensitive is the synaptic drive at one neuron to the
precise timing of spikes emitted by the other? To formulate this explicitly, suppose that a single spike time
$t_A^m$ of neuron $A$ is perturbed by a small amount $\epsilon$. The question then is how does this affect the
synaptic current at neuron $B$? Addressing this necessitates the distributional derivative formalism we developed in Section \ref{sec:spike.diffs}. We will first look at how we do this for the case of an approximated single spike, and then extend the result to a spike train exactly by making use of the distributional derivative. 

From Equation \ref{eq:synaptic_drive_B}, the contribution of spike $m$ to the
synaptic drive at $B$ is
\begin{equation} \nonumber
I_m(t) := K\big(t - t_A^m - \tau_{AB}\big)
\end{equation}
If the spike time shifts from $t_A^m$ to $t_A^m + \epsilon$, this contribution becomes
\begin{equation}
I_m^\epsilon(t) :=K\big(t - (t_A^m + \epsilon) - \tau_{AB}\big)
= K\big(t - t_A^m - \tau_{AB} - \epsilon\big)
\end{equation}
Define the shorthand
\begin{equation}\nonumber
\sigma := t - t_A^m - \tau_{AB}
\end{equation}
so that the unperturbed contribution is $K(\sigma)$ and the perturbed contribution is
$K(\sigma - \epsilon)$.

Because $K(t)$ is smooth for $t > 0$, the bi-exponential kernel is infinitely
differentiable on its support and we can expand $K(\sigma - \epsilon)$ as a Taylor series
around $\sigma$, written as
\begin{equation}\nonumber
K(\sigma - \epsilon) = K(\sigma) - \epsilon \, K'(\sigma)
+ \frac{\epsilon^2}{2} K''(\sigma) - \cdots
\end{equation}
This is the standard Taylor expansion of a smooth function $f$ about a point $a$,
applied to $f = K$, $a = \sigma$, and the displacement $= -\epsilon$
\begin{equation}\nonumber
f(a + h) = f(a) + h\,f'(a) + \frac{h^2}{2}f''(a) + \cdots
\end{equation}
with $h = -\epsilon$.

For small $\epsilon$, we only need to keep the first two terms and can discard all terms of order
$\epsilon^2$ and higher. In other words, when
\begin{equation}\nonumber
K(\sigma - \epsilon) \approx K(\sigma) - \epsilon \, K'(\sigma)
\end{equation}
This approximation is valid whenever $|\epsilon|$ is small enough that the neglected
terms are negligible, i.e., whenever
\begin{equation}\nonumber
\frac{\epsilon^2}{2} |K''(\sigma)| \ll \epsilon \, |K'(\sigma)|
\end{equation}
In practice, this
holds whenever the timing perturbation $\epsilon$ is small relative to the temporal
scale on which the kernel changes shape, which is set by the synaptic time constants
$\tau_r$ and $\tau_d$.

The change in the synaptic drive due to the perturbation is the difference between the
perturbed and unperturbed contributions:
\begin{align} 
\Delta I_m(t) &:= I_m^\epsilon(t) - I_m(t) \nonumber \\
&= K(\sigma - \epsilon) - K(\sigma) \nonumber \\
&\approx \big[K(\sigma) - \epsilon \, K'(\sigma)\big] - K(\sigma) \nonumber \\
&= -\epsilon \, K'(\sigma) \nonumber
\end{align}
Substituting back $\sigma = t - t_A^m - \tau_{AB}$ gives
\begin{equation}
\Delta I_{A \to B}(t) \approx -\epsilon \, K'\big(t - t_A^m - \tau_{AB}\big)
\end{equation}
where $K'$ denotes the ordinary derivative of the kernel. The change in synaptic drive at neuron $B$, caused by a small
shift $\epsilon$ in a spike time of neuron $A$, is proportional to the derivative of the
postsynaptic kernel evaluated at the arrival time. The minus sign has an explicit physical
interpretation: It delays the spike (positive $\epsilon$) and shifts the kernel to the right on the
time axis, which at a fixed observation time $t$ means the kernel is being sampled at
an earlier point in its waveform. This reduces its value if the kernel is in its rising
phase. $K'(\sigma)$, i.e., the slope of the kernel at the moment the
spike's effect is felt, determines whether the downstream current is sensitive or
insensitive to small timing perturbations. Where the kernel rises steeply
($|K'|$ large), timing precision matters; where it decays slowly ($|K'|$ small),
it does not.

Now consider the connection to the distributional derivative. The distributional derivative we developed in Section \ref{sec:spike.diffs} extends this observation
from a single-spike approximation to an exact closed-form result for the entire spike
train. Recall that the derivative of a spike train distribution acts on a test function
$\varphi$ as
\begin{equation}\nonumber
\langle s_A', \varphi \rangle = -\sum_m \varphi'(t_A^m)
\end{equation}

If we take the test function $\varphi$ to be the synaptic kernel $K$, evaluated as a
function of the presynaptic spike time for a fixed observation time $t$ and latency
$\tau_{AB}$, then by the chain rule
$\frac{\partial}{\partial t_A^m} K(t - t_A^m - \tau_{AB}) = -K'(t - t_A^m - \tau_{AB})$,
and the distributional derivative gives
\begin{equation}
-\sum_m \frac{\partial}{\partial t_A^m} K\big(t - t_A^m - \tau_{AB}\big)
= \sum_m K'\big(t - t_A^m - \tau_{AB}\big)
\label{eq:dist_deriv_kernel}
\end{equation}
Equation \ref{eq:dist_deriv_kernel} is the total timing sensitivity of the synaptic
drive at neuron $B$ to perturbations in the spike times of neuron $A$, summed across
all spikes. Unlike the Taylor expansion, this is \textit{not} an approximation, it is an
exact calculation and consequence of the distributional framework applied to the full spike train.

Importantly, this computation is not well-defined without the distributional framework. The spike
train $s_A(t)$ has no pointwise derivative; it is not a function in the classical
sense. The expression $\langle s_A', \varphi \rangle$ only has meaning because $s_A'$ is
defined as a distribution (Section \ref{sec:spike.diffs}), acting on smooth test functions via the
integration by parts identity. The kernel $K(t)$, being smooth for $t > 0$, serves as
the test function in this context, which is why the computation produces a well-defined
result.

\subsection{Causal admissibility via distributional support} \label{sec:causal.admin}

The last question we address is which arriving spikes can actually influence the receiving
neuron at all?

After neuron $B$ emits a spike at time $t_B^n$, it enters a refractory period of
duration $R_B$. During the interval $(t_B^n, \, t_B^n + R_B)$, neuron $B$ cannot respond
to any input, regardless of stimulus strength. We define the refractory window of neuron
$B$ following its $n$-th spike as
\begin{equation}
\mathcal{R}_B^n := (t_B^n, \, t_B^n + R_B)
\end{equation}

Now consider a spike from neuron $A$ initiated at time $t_A^m$. This spike arrives at
neuron $B$ at time $t_A^m + \tau_{AB}$. Any spike is \textit{causally admissible}, in the sense that
it can potentially influence neuron $B$ by contributing to subthreshold summations towards the threshold potential, if and only if the arrival time falls
outside the refractory window:
\begin{equation}
t_A^m + \tau_{AB} \notin \bigcup_n \mathcal{R}_B^n
\end{equation}
If the arrival time falls inside the refractory window, the spike is causally excluded, meaning
it does not contribute to the state of neuron $B$, regardless of the synaptic weight
or kernel amplitude.

This condition can be stated precisely and explicitly in the language of distributional support, which is necessary for exact calculations of the refractory constraint.
The delayed spike train $s_A(t - \tau_{AB})$ has support at the arrival times
$\{t_A^m + \tau_{AB}\}$ (Equation \ref{eq:delayed_train}). The refractory state of neuron
$B$ defines a set of times $\bigcup_n \mathcal{R}_B^n$ during which $B$ is blind to an input.
A spike from $A$ is admissible at $B$ if and only if its support point (arrival time)
lies outside the refractory set:
\begin{equation}
\mathrm{supp}\big(\delta(t - t_A^m - \tau_{AB})\big) \cap \bigcup_n \mathcal{R}_B^n
= \emptyset
\label{eq:support_admissibility}
\end{equation}
When this condition holds, the spike contributes to the synaptic drive as in Equation
\ref{eq:synaptic_drive_B}. When it fails, the spike is excluded entirely. This reflects a necessary condition for contributing to synaptic drive (Section \ref{sec:synaptic.drive}) in the first place, and any subsequent analysis of spike time sensitivity (Section \ref{sec:spke.sensitivity}).

This is a distributional statement about the relationship between the support of
individual spike events and the temporal structure of the receiving neuron's refractory
state. It cannot be formulated in terms of pointwise function values because the spike
train has no pointwise values, it is a distribution defined property only through its action
on test functions. The concept of distributional support (Section \ref{sec:test.functions.i}) is what makes
this condition precise. A spike event is `located' at a specific time in the sense
that $\mathrm{supp}(\delta(t - a)) = \{a\}$, and the question of whether it can influence
a downstream neuron reduces to a set-theoretic question about the relationship between
this support point and the refractory intervals.

\section{Discussion: What the distributional framework enables}
The distributional representation of spike trains we develop from first principles in this paper is a powerful mathematically formal and exact framework for manipulating and understanding the dynamic structure and effects of action potentials as discrete (digital) signaling events interacting with (analog) continuous physiological and biophysical variables. The two-neuron circuit example in Section \ref{sec:2.neuron.circuit} explored the application of the framework on three concrete analytic results that can either only be approximated by continous method techniques, or cannot be modeled or computed at all. 

First, it allowed us to compute the \textbf{exact synaptic drive} (Section \ref{sec:synaptic.drive}) for each neuron in the circuit. The convolution of a spike train
distribution with a synaptic kernel gives the exact, continuous-time synaptic current
as a closed-form sum over copies of a spike-triggered kernel. There is no discretization or rate
approximation that is required.

Second, we determined the \textbf{exact timing sensitivity and dependency} on arriving spikes (Section \ref{sec:spke.sensitivity}). The distributional derivative
of a spike train, composed with the synaptic kernel, yields the exact sensitivity of
the postsynaptic current to perturbations in individual spike times. This computation
is well-defined only because the spike train derivative exists in the distributional
sense.

Lastly, we defined and developed a condition for \textbf{precise causal admissibility of inputs} (Section \ref{sec:causal.admin}). The distributional support
of individual spike events provides a precise condition for whether an arriving spike
can influence the receiving neuron, given the geometry of the connection, signaling event latencies, and the receiving neuron's refractory state. This is a
set-theoretic condition on distributional support (\textit{c.f.} \cite{sila2019}), not a pointwise function evaluation.

None of these results require smoothing the spike train into a firing rate, discretizing
time into bins, or treating the Dirac delta as an informal shorthand. They follow
directly by representing spike trains as what they mathematically are,
distributions in the sense of Schwartz, and applying the convolution,
differentiation, and support operations as defined by distribution theory.

This example involved only two neurons, but the analysis generalizes to
networks. In a network of $N$ neurons with heterogeneous delays $\{\tau_{ij}\}$ and
refractory durations $\{R_j\}$, the distributional framework provides the exact synaptic drive via
convolution, the timing sensitivity via the distributional derivative, and the causal
admissibility condition via support. The interplay among these quantities, for
example, how refractory gating selectively excludes spikes arriving along
short-delay pathways while admitting those arriving along long-delay pathways, gives
rise to timing-dependent structure in effective connectivity that emerges purely from
the interaction of physical constraints with event timing. The distributional framework
developed in this paper provides the mathematical foundation for such analyses.

The representational and analytic power of the distributional framework is not limited to circuits or networks of neurons, but naturally also accommodates biophysical variability that is often treated as
extraneous to spike-train models. For example, action potential waveforms vary significantly across
neuronal compartments. Somatic spikes are typically broader than axonal spikes, and
presynaptic terminal waveforms can differ from both depending on the local ion channel
composition or recent activity history \cite{foust2011somatic}. These waveform
differences can have real functional consequences. Broader presynaptic action potentials enhance
calcium influx and increase neurotransmitter release, directly modulating synaptic
strength. In the distributional framework, this variability is not modeled through the spike
train directly, since the spike train only encodes the record of event times, but rather through the postsynaptic
kernel $K(t)$. Different presynaptic waveform conditions correspond to different
kernels, and the convolution $(K * s_i)(t) = \sum_k K(t - t_i^k)$ produces
correspondingly different synaptic currents. The kernel can be allowed to vary across
synapses, axonal compartments, or even dynamically as a function of recent
spiking history, such as short-term facilitation or depression. This approach preserves the distinct
separation between event timing on the one hand and synaptic effects on the other,
while naturally incorporating a layer of biophysical realism that is often assumed to
require waveform-level modeling considerations.

The framework presented here establishes the mathematical foundation. Its application to the analysis of timing-dependent computation in neuronal dynamics and neural circuits is the subject of ongoing work.

\bibliographystyle{unsrt} 
\bibliography{refs}

\end{document}